\documentclass[useAMS,usenatbib]{mn2e}

\usepackage{amsmath}

\usepackage{epsfig}
\usepackage{epstopdf}
\usepackage{lscape} 
\usepackage{natbib}
\usepackage{tabularx}
\usepackage{multirow}
\usepackage{amssymb}
\def\gtrsim{\mathrel{\hbox{\rlap{\hbox{\lower4pt\hbox{$\sim$}}}\hbox{$>$}}}}

\title{Non-Cosmological FRBs from Young Supernova Remnant Pulsars}
\author[Connor et al.]{
Liam Connor$^{1,2,3}$\thanks{E-mail:\ connor@astro.utoronto.ca}
Jonathan Sievers $^{4, 5}$\thanks{E-mail:\ jonathan.sievers@gmail.com}
Ue-Li Pen $^{1, 6, 7}$\thanks{E-mail:\ pen@cita.utoronto.ca}
\\
$^1$ Canadian Institute for Theoretical Astrophysics, University of Toronto, M5S 3H8 Ontario, Canada
\\
$^2$ Department of Astronomy and Astrophysics, University of Toronto, 
M5S 3H8 Ontario, Canada
\\
$^3$ Dunlap Institute for Astronomy and Astrophysics, University of Toronto,
Toronto, ON M5S 3H4, Canada
\\
$^4$ Astrophysics and Cosmology Research Unit, School of Chemistry and Physics, University of KwaZulu-Natal, Durban, South Africa
\\
$^5$ National Institute for Theoretical Physics, KwaZulu-Natal, South Africa
\\
$^6$ Canadian Institute for Advanced Research, Program in Cosmology
and Gravitation
\\
$^7$ Perimeter Institute for Theoretical Physics, 31 Caroline St. N., Waterloo, ON, N2L 2Y5, Canada
}

\begin{document}
\date{\today}
\pagerange{\pageref{firstpage}--\pageref{lastpage}} 
\pubyear{2015}
\maketitle
\label{firstpage}

\begin{abstract}
We propose a new extragalactic but non-cosmological explanation for fast radio bursts 
(FRBs) based on
very young pulsars in supernova remnants. Within a few hundred years of a 
core-collapse supernova the ejecta 
is confined within $\sim$1 pc, providing a high enough column density of free electrons 
for the observed 375-1600 pc cm$^{-3}$ of dispersion measure (DM). By extrapolating a Crab-like pulsar to 
its infancy in an environment like that of SN 1987A, 
we hypothesize such an object could emit supergiant pulses sporadically which 
would be bright enough to be seen at a few hundred megaparsecs. We hypothesize that
such supergiant pulses would preferentially occur early in the pulsar's life when the 
free electron density is still high, which is why we do not see large numbers of 
moderate DM FRBs ($\lesssim300$ pc cm$^{-3}$). 
In this scenario Faraday
rotation at the source gives rotation measures (RMs) much larger than the expected
cosmological contribution.  If the emission were pulsar-like, then the polarization 
vector could swing over the duration of the burst, which is not expected from 
non-rotating objects.
In this model, the scattering,
large DM, and commensurate RM all come from one place which is not the case for the cosmological
interpretation. The model also provides
testable predictions of the flux distribution and repeat rate of FRBs, and could be further
verified by spatial coincidence with optical supernovae of the past several decades and
cross-correlation with nearby galaxy maps. 
\end{abstract}
\begin{keywords}
FRB, supernova remnants, pulsar, giant pulse
\end{keywords}


\newcommand{\be}{\begin{eqnarray}}
\newcommand{\ee}{\end{eqnarray}}
\newcommand{\beq}{\begin{equation}}
\newcommand{\eeq}{\end{equation}}

\section{Introduction}
The mystery of fast radio bursts (FRBs) has garnered
substantial interest from the radio community.
High-energy astrophysicists have tried to model their burst source, 
observers would like to measure a large population of them, and cosmologists
hope to use them as a probe of the intergalactic medium (IGM). However their relative scarcity 
(only $\sim$ dozen have been observed so far) and their apparent 
transient nature mean that we still do not know their position on the sky
to better than a few arcminutes, and their radial position could be anything
from terrestrial to cosmological \citep{2014ApJ...797...70K}.

These objects are
highly dispersed, with dispersion measures (DMs) ($\sim 375$-1600 pc cm$^{-3}$) far exceeding
the expected contribution from our own Galaxy's interstellar medium (ISM) (10-100 pc cm$^{-3}$) and leading to the
interpretation that FRBs are cosmological \citep{2007Sci...318..777L, 2013Sci...341...53T}. 
Various emission mechanisms have been proposed 
at a wide range of source locations, 
including merging white dwarfs \citep{2013ApJ...776L..39K}
and neutron stars \citep{2013PASJ...65L..12T}, supergiant pulses 
from extragalactic neutron stars \citep{2015arXiv150100753C},
blitzars \citep{2014A&A...562A.137F}, 
magnetars \citep{2007arXiv0710.2006P, 2014MNRAS.442L...9L, 2015arXiv150101341P}, 
and flaring Galactic stars \citep{2014MNRAS.439L..46L}. 
Though presently there are more theoretical models for FRBs than actual 
sources discovered, constraints on such theories are rapidly emerging. 
This is due to recent polarization data, 
multifrequency coverage, and their being observed by several telescopes
at various locations on the sky \citep{2014ApJ...780L...2B, 2015MNRAS.447..246P}. 

On top of event rates ($\sim$10$^4$ per day per sky) 
and high DMs, explanations of FRBs must now
account for temporal scattering, and polarization states.  They should
predict or explain Faraday rotation and time dependence of linear polarization.
The rotation measure (RM) of our Galaxy has been mapped, and the
intergalactic RM is constrained to be less than 7 rad m$^{-2}$
\citep{2015A&A...575A.118O}. 
The observed temporal scattering is problematic for a IGM interpretation, due 
to the unrealistically small length scales required in the IGM 
for $\sim$ms scattering \citep{2014ApJ...785L..26L}. 

In this letter we propose a new non-cosmological but extragalactic
solution to the FRB problem: supergiant pulses from newly formed pulsars in 
supernova remnants (SNRs). The dense ionized environment of the SNR
can provide 300-2000 pc cm$^{-3}$ of dispersion if the pulses are observed 
within $\sim100$ years of the core-collapse supernova. In our picture the 
large DM and scattering all come from the same place, and generically
accounts for substantial Faraday Rotation and polarization angle
swings. These features were included to account for recent
polarization measurements of a new FRB (Masui et al. 2015, submitted),
which may exhibit a polarization vector swing and whose RM is $\sim2\times10^2$ rad m$^{-2}$.
These are not expected in a cosmological interpretation of the DM.


\section{Supernova Remants}
Of order 10$^{51}$ ergs of kinetic energy is released during a supernova, a 
fraction of which is converted into thermal 
energy after shock heating of the 
ejecta plasma. Though the shock-heated ejecta atoms 
are fully ionized after the explosion, the density is high enough that
ionized atoms can soon recombine.
This phase of low-ionization comes to an end when the remnant expands 
into the surrounding ISM, causing a reverse shock wave that reionizes the ejecta.
Though this is the basic narrative, observations \citep{2014ApJ...796...82Z} 
as well as simulations \citep{2014ApJ...794..174P}
of SN 1987a have shown the morphological and ionization properties of SNRs
in the decades and centuries after the explosion are nuanced and 
difficult to model.
That said, in general the expanding shell left behind 
should be able to provide enough free electrons
along the line-of-sight for unusually large dispersion measures. If we 
assume a toy model in which a sphere expands at $v_{\rm ej}$, 
then the radius R(t) $\approx v_{\rm ej} t$. Therefore the DM we expect can be 
calculated as,

\begin{equation}
\textup{DM} \approx  \frac{\textup{x}_e \textup{M}_{\rm ej}}{m_p \frac{4\pi}{3} v_{\rm ej}^2 t^2}
\end{equation}

\noindent where x$_e$ is the ionization fraction, 
M$_{\rm ej}$ is the ejecta mass, and $m_p$ 
is the mass of a proton. Assuming $\sim$10 M$_{\odot}$ of material 
is ejected at $v_{\rm ej}\sim 3-8\times10^3$ km/s and an ionization fraction of 
$\sim20\% $, the dispersion measure goes from several 
thousand pc cm$^{-3}$ immediately
after the reverse-shock ionization, to several hundred pc cm$^{-3}$ after 50-100 years 
\cite{2014ApJ...796...82Z}. We point out that while the difference between between 
a sphere of HII, which we have assumed, and a thin shell makes a small difference
for DM, it could have a large effect of plasma frequency -- something we discuss
in section \ref{sec-predictions}.
In the context of SNR 1987a, \cite{2014ApJ...796...82Z} found that a possible pulsar 
could have DMs between 100-6000 pc cm$^{-3}$, after $\sim 25$ years, though 
no compact object has yet been observed in that remnant.

Another potentially important feature of the SNR environment is its magnetic
field. 
The exact magnitude of any detection of Faraday rotation has implications for the possible source location. For
instance in the circumnuclear picture, one would expect RMs 
$\sim10^{3-5}$ rad m$^{-2}$ \citep{2015arXiv150101341P}, similar to that of the Milky Way's
Galactic center magnetar J1745-29. In the cosmological scenario, if the Faraday 
rotation
came from the same place as the DM - namely the intergalactic medium -
then we would only expect a few rad m$^{-2}$ of RM \citep{2015A&A...575A.118O}. 

The Faraday effect rotates the polarization vector
by an angle $\phi = $RM$\, \lambda^2$, where

\begin{equation}
\textup{RM} = \frac{e^3}{2\pi m^2 c^4} \int_0^{L} n_e(l) B_\parallel (l) dl.
\end{equation}

We can therefore make a rough estimate of the rotation measure of a remnant 
pulsar with dispersion measure DM. Using 
\cite{2014ira..book.....B} we get,

\begin{equation}
\textup{RM} \approx 0.81\,\textup{rad}  \, \textup{m}^{-2} \, \times \frac{\left < B_{\parallel} \right >}
{1 \mu \rm G} \cdot \frac{\textup{DM}}{1 \textup{pc}\,\textup{cm}^{-3}} .
\end{equation}

Though there is a large uncertainty in evolution of the magnetic field strength and added
uncertainty in $\langle B_{\parallel} \rangle$ given $B_{\parallel}$ is not necessarily positive, 
typical values in our Galaxy are 0.2 - 1$\mu \rm G$. For instance the Crab and Vela have 
$ \sim 0.92 \mu \rm G$ and $\sim 0.56 \mu \rm G$, respectively. 
This gives RMs between $\sim 80-1200$
rad m$^{-2}$ for a SNR pulsar with FRB-like DMs, which is consistent with Masui et al. (2015, 
submitted).

\subsection{Event Rates}
\label{rates}

\begin{figure}
\label{FIG-RATE}
  \centering
   \includegraphics[width=0.5\textwidth]{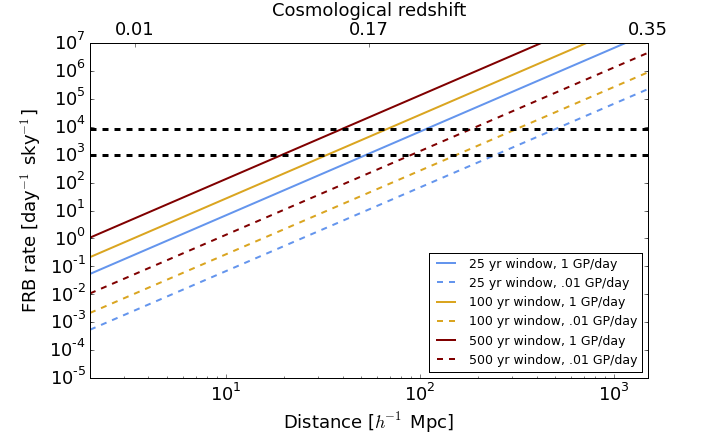}
   \caption{Daily FRB rate per sky based on local core-collapse supernova 
   event rate, plotted against distance.
   We assume early in the pulsar's life there is a window, either 
   25, 100, or 500 years when the SNR can provide a large enough electron 
   column density to explain the high DMs of the observed bursts. We also
   include a rate of giant pulses of either one per day or one per hundred
   days. We have assumed 20$\%$ of core-collapse supernovae leave behind
   a visible pulsar.
   The horizontal black lines are the $99\%$ confidence bounds for the FRB rate
   found by \protect\cite{2015arXiv150500834R}.}
\end{figure}

The daily FRB rate has been estimated at $3.3^{+5.0}_{-2.5}\times10^3$ sky$^{-1}$ 
\citep{2015arXiv150500834R}. If we start from the local core-collapse supernova
event rate, $\Gamma_{\rm CC}$, and include objects out to some distance $d_{\rm max}$,
we expect the following daily FRB rate, 

\begin{equation}
\Gamma_{\rm FRB} \sim  \frac{4}{3} \pi d_{\rm max}^3 \times \Gamma_{\rm CC} \times
 \eta \, \tau_{\rm ion} \gamma_{\rm GP}
\end{equation}

\noindent where $\tau_{\rm ion}$ is the window in years when the SNR is sufficiently
dense and ionized to provide the observed DMs, $\gamma_{\rm GP}$
is the daily rate of giant pulses above $\sim10^{36}$ ergs, and $\eta$
is the number of core-collapse supernovae that leave behind a visible pulsar. 
From \cite{2014ApJ...792..135T} we know  
$\Gamma_{\rm CC}\sim3 \times 10^{-4}$ day$^{-1}$ ($h^{-1}$Mpc)$^{-3}$,
so if we take $d_{\rm max}$ to be 100 $h^{-1}$Mpc and $\tau_{\rm ion}\sim100$ years,
we require one giant pulse every 10-20 days, assuming one fifth of this SNe population
leaves behind a visible pulsar.
In Fig. \ref{FIG-RATE} 
we show the event rate as a function of distance, varying two parameters: the 
effective high-DM window and the rate of giant pulses. 

From Fig. \ref{FIG-RATE} we can see even in our most conservative estimate, when
the SNR only has a 25 year window and emits giant pulses once every
100 days, the volume necessary for the highest daily FRB rate is still non-cosmological.
By this we mean the DM contribution from the IGM is less than $\sim 200$ pc cm$^{-3}$.
If the SNR FRBs are within a hundred $h^{-1}$Mpc then DM$_{\rm IGM}$ is less than 
$\sim 10 \%$  of the total dispersion of a typical burst.

If FRBs really are giant pulses then they 
should repeat stochastically, and while none of the radio follow-ups for 
observed sources has seen an FRB repeat, this could be because they
have not observed for long enough. We point out that FRB 140514, 
the first burst observed in real-time, was found during a follow up observation
of FRB 110220 and the two were found within a beam-width of one another.
FRB 140514 had a lower DM than 110220 by 380 cm pc$^{-3}$, 
and though \cite{2015MNRAS.447..246P} show that it was not very unlikely that one would find a 
new FRB given their integration time, if it were the same source 
our model could explain the discrepancy. Indeed, a reanalysis by 
\cite{2015arXiv150701002M} found that the two bursts were far more likely
to be the same repeating source than had been previously claimed.
Given FRB 110220 would have been emitted over three years earlier, we 
would expect its DM generically to be higher, but the amount would 
depend on the inner structure of the SNR and its expansion speed. 
We discuss repetition further in section \ref{sec-predictions}.

\subsection{Young SNR Pulsars}

About a dozen pulsars in our Galaxy are known to emit extremely energetic,
short duration radio pulses which can be many orders of magnitude 
brighter than the pulsar's regular emission. Some of these objects exhibit 
a rare tail of \textit{supergiant} pulses, whose brightness temperatures 
exceed the Planck temperature, $\gtrsim10^{32}$ K \citep{2004ApJ...612..375C},
which we will take as a working definition of \textit{supergiant}. 
Indeed the largest
known brightness temperature, T$_{b}$, in the universe came from a giant pulse from the Crab,
 with T$_{b}\sim2\times10^{41}$ K \citep{2014ApJ...792..135T}. Though there
are only $\sim100$ hours of published giant pulse data from the Crab, it is known
that the supergiant pulse tail does not obey the standard power-law fall off
in amplitude \citep{2012ApJ...760...64M}.

Given the relatively high frequency of core-collapse supernovae 
in the local universe, the young 
rapidly rotating pulsars such events leave behind could emit giant 
pulses bright enough to be observed at hundreds of megaparsecs. 
These supergiant pulses would require $10^{36 - 37}$ ergs of output,
assuming and observed flux density of 0.3-5 Jy and $\sim500$ 
MHz of bandwidth over 1 ms. Though this is $\sim$ billions of times
brighter than an average pulse, it is negligible compared to a 
pulsar's total rotational energy, E$_{\rm rot} \sim 10^{49-50}$ ergs and
even the pulsar's spin-down luminosity. We
also point out that given its relative proximity, this model requires
a couple orders of magnitude less energy than cosmological FRBs,
located beyond a Gpc. 


\begin{table*}
\label{TAB-1}
\begin{tabularx}{1.08\textwidth}{@{\extracolsep{\fill}}|lccccccc|}
\hline
\multicolumn{1}{|c}{\textbf{Location}}                                                                                            & \textbf{Model}                                              & \textbf{\begin{tabular}[c]{@{}c@{}}Galactic \\ scintillation\end{tabular}} & \textbf{\begin{tabular}[c]{@{}c@{}}Faraday \\ rotation\end{tabular}} & \textbf{$\mathbf{\frac{dlnN_{\rm FRB}}{dlnS_{\nu}}}$}                                      & \multicolumn{1}{l}{\textbf{Counterpart}}                                    & \textbf{\begin{tabular}[c]{@{}c@{}}DM range\\ (pc cm$^{-3}$)\end{tabular}} & \textbf{\begin{tabular}[c]{@{}c@{}}Pol angle \\ swing\end{tabular}} \\ \hline
\multicolumn{1}{|l|}{\multirow{4}{*}{\begin{tabular}[c]{@{}l@{}}Cosmological \\ ($\gtrsim 1 h^{-1}$Gpc)\end{tabular}}}             & Blitzars                                                    & $\times$                                                                         & $\lesssim 7$ rad m$^{-2}$                                               & ?                                                                                      & \begin{tabular}[c]{@{}c@{}}gravitational \\ waves\end{tabular}              & 300-2500                                                                & $\times$                                                                  \\
\multicolumn{1}{|l|}{}                                                                                                            & Merging COs                                                 & $\times$                                                                         & $\lesssim 7$ rad m$^{-2}$                                      & ?                                                                                      & \begin{tabular}[c]{@{}c@{}}type Ia SNe,\\  X-ray, $\gamma$-ray\end{tabular} & 300-2500                                                                & $\times$                                                                  \\
\multicolumn{1}{|l|}{}                                                                                                            & Primordial BHs                                              & $\times$                                                                         & $\lesssim 7$ rad m$^{-2}$                                               & ?                                                                             & $\sim$TeV                                                                   & 300-2500                                                                & $\times$                                                                  \\
\multicolumn{1}{|l|}{}                                                                                                            & Magnetar flare                                              & $\times$                                                                         & $\lesssim 7$ rad m$^{-2}$                                               & ?                                                                                      & \begin{tabular}[c]{@{}c@{}}$\sim$ms TeV \\ burst\end{tabular}               & 300-2500                                                                & $\checkmark$                                                        \\ \cline{1-1}
\multicolumn{1}{|l|}{\multirow{3}{*}{\begin{tabular}[c]{@{}l@{}}Extragalactic, local \\ ($\lesssim$200$ h^{-1}$Mpc)\end{tabular}}} & Edge-on disk                                                & $\checkmark$                                                               & 50-500 rad m$^{-2}$                                                     & -3/2                                                                                   & ?                                                                           & 10-2000                                                                 & ?                                                                   \\
\multicolumn{1}{|l|}{}                                                                                                            & \begin{tabular}[c]{@{}c@{}}Nuclear \\ magnetar\end{tabular} & $\checkmark$                                                               & 10$^{3-5}$ rad m$^{-2}$                                                 & -3/2                                                                                   & none                                                                        & 10-3000                                                                 & $\checkmark$                                                        \\
\multicolumn{1}{|l|}{}                                                                                                            & SNR pulsar                                                  & $\checkmark$                                                               & 20-$10^3$ rad m$^{-2}$                                                  & -3/2                                                                                   & \begin{tabular}[c]{@{}c@{}}archival CC \\ SNe or \\ nearby galaxy \end{tabular}                  & 10$^2$-10$^4$                                                           & $\checkmark$                                                        \\ \cline{1-1}
\multicolumn{1}{|l|}{Galactic ($\lesssim 100$ kpc)}                                                                                & flaring MS stars                                            & $\checkmark$                                                               & RM$_{\rm gal}$                                                          & -3/2                                                                                   & \begin{tabular}[c]{@{}c@{}}main sequence \\ star\end{tabular}               & $\gtrsim$ 300                                                           & $\times$                                                                  \\ \cline{1-1}
\multicolumn{1}{|l|}{Terrestrial ($\lesssim 10^5$ km)}                                                                             & RFI                                                         & $\times$                                                                         & $\lesssim$ RM$_{\rm ion}$                                                          & $\left\{\begin{matrix}-1/2 \,\, if \,\, 2D \\ -3/2 \,\, if \,\, 3D\end{matrix}\right.$ & none                                                                        & ?                                                                       & $\times$                                                                  \\ \hline

\end{tabularx}
\caption{This table summarizes a number of FRB models by classifying them as cosmological, 
extragalactic but non-cosmological, Galactic, and terrestrial. 
The seven columns are potential observables of FRBs and each
 row gives their consequence for a given model 
 (Blitzars \protect\citep{2014A&A...562A.137F}, 
 compact object mergers \protect\citep{2012ApJ...760...64M, 2013PASJ...65L..12T},
 exploding primordial blackholes \protect\citep{2014PhRvD..90l7503B}, 
bursts from magnetars \protect\citep{2014MNRAS.442L...9L}, 
edge-on disk galaxies \protect\citep{2015arXiv150400200X}, 
circumnuclear magnetars \protect\citep{2015arXiv150101341P}, 
 supernova remnant pulsars, stellar flares \protect\citep{2014MNRAS.439L..46L}, and terrestrial RFI 
 \protect\citep{2015arXiv150305245H}.). For the latter, we subdivide the RFI into planar RFI (2D) coming
 from the earth's surface, and 3D RFI coming from objects like satellites. 
 Since scintillation
only affects unresolved images, cosmological sources that are not scattered near the source
will not scintillate in our Galaxy, while non-cosmological sources whose screens are intrinsic will. 
For Faraday rotation and scintillation
 we assume 
the RM and SM comes from the same place as the DM, e.g. the IGM for cosmological sources, though such models 
could introduce a more local Faraday effect or a scattering screen.
Even though
all models have to explain the observed 375-1600 pc cm$^{-3}$, some models predict a wider 
range of DM. For instance, in the circumnuclear magnetar or edge-on disk disk scenarios there 
ought to be bursts at relatively low DM that simply have not been identified as FRBs. In our supernova 
remnant model DMs should be very large early in the pulsar's life, though this window is short and 
therefore such high DM bursts would be rare.}
\end{table*}
2015MNRAS.447..246P

The polarization properties of giant pulses are also consistent with those
of observed FRBs. Giant pulses are known to be highly polarized, 
switching between strong Stokes V and purely linearly polarized states often
in an unpredictable way.
The only published FRB prior to Masui et al. (2015, submitted), 
with full-pol information was FRB 140514 and was 
found to have $\sim20\%$ circular polarization and no detectable 
linear polarization \citep{2015MNRAS.447..246P}. If FRBs were coming
from a pulsar-like emission mechanism, one might see nearly pure Stokes V
or linear-pol states. Another consequence of pulsar-like emission 
is that FRBs could exhibit polarization angle
swings over the burst duration, which may have been observed 
by Masui et al. (2015, submitted). 
Unfortunately, to date all other published FRBs
were detected with systems that recorded only Stokes I. 

\section{Predictions}
\label{sec-predictions}

In table 1 we summarize the observational consequences
of ours and several other models as best we can. As one might expect,
the most striking differences in predictions has to do with the distance of FRBs,
for example the cosmological FRB models differ mainly in their expected
counterpart and not much else. 

The young SNR pulsar model makes several predictions that will
be addressed with more data, particularly with full polarization 
observations and large field-of-view surveys. 
The latter will provide a large sample of FRBs whose flux and DM statistics
 can give us information about their location. Since in the SNR FRB picture
most of the DM is intrinsic, the sources do not need to be at cosmological 
distances. This means the flux distribution is given by a Euclidean universe
that is only weakly dependent on DM, N$(>$S)$ \propto \rm S^{-3/2}$, assuming
the bursts are standard candle-like.
Wide-field surveys
like CHIME \citep{2014SPIE.9145E..22B} (whose FRB backend will 
observe steadily for several years), 
UTMOST\footnote{http://www.caastro.org/news/2014-utmost},
 or HIRAX could observe as many as $\sim10^{3-4}$
per year, which would allow for detailed population statistics. An instrument like 
CHIME will not only give us large numbers of DMs and fluxes, but will also allow 
us to measure various polarization properties and frequency scintillation. 

Since we have proposed that FRBs come from young pulsars 
in SNRs, it is possible that the corresponding 
supernova was observed in recent decades in the optical. If the pulsars
were younger than $\sim$60 years old they could be localised at the 
$\sim$arcsecond level
and matched against catalogued type II supernovae, though we would need a 
large sample of FRBs given the incompleteness of recorded supernovae. With
current data the location of FRBs has been too poorly constrained to say anything 
meaningful about overlap with historic SNe or coincident galaxies; 
out to $\sim$150 $h^{-1}$Mpc 
there are a number of galaxies in a Parkes beam and therefore one would 
expect as many supernovae anyway in the last century, even though it would unlikely 
have been observed. However better localisation or a cross-correlation between
a large sample of FRBs and nearby galaxies could help support the non-cosmological
extragalactic FRB hypothesis.


We also point out that while FRBs seem not to repeat regularly, 
it is not known that they never repeat. Though the statistics 
of giant pulses from local pulsars are mostly Poisson \citep{1999ApJ...517..460S},
it is possible that 
the supergiant pulses we require from very young SNR pulsars are not. If their 
statistics were of a Poisson process then there are already limits on the repeat rate, 
given the $\sim100$ hours of follow up, however if their statistics were more like
earthquakes, the brightest pulses could burst intermittently and turn off
for extended periods. It is possible that FRBs could repeat every 5-500 days. 
If they were to repeat,
it is possible that their DMs, RMs, and scattering properties could 
change noticeably on months/years timescales. Unlike standard pulsars 
whose RMs and DMs are constant to a couple decimal places, young 
SNR pulsars like the Crab and Vela have shown significant - and sometimes
correlated - variation in such properties \citep{1988A&A...202..166R}.
As discussed in section \ref{rates}, FRB 140514 had a DM that was several hundred
pc cm$^{-3}$ smaller than FRB 110220 and the two were found within a beam-width 
of one another. Though this could have just been a spatial coincidence of two
separate objects, our SNR FRB model 
could account for such a change in DM while other models (cosmological, edge-on 
galaxy, etc.) cannot. 

We also predict that such repeated 
bursts could have vastly different polarization states, similar to the giant 
pulses from pulsars in the Galaxy. Another consequence of polarized 
pulsar-like emission would be a polarization angle swing. Given the FRBs
would be rotating, the angle of the linear polarization vector could 
change throughout the pulse -- a phenomenon that is seen in many 
galactic pulsars, often repeatably \citep{2006ApJ...645.1421B}. Therefore models 
that explain FRBs as rapidly rotating compact objects could predict a swing in the 
polarization angle throughout the burst.

Depending on the relationship between the giant pulse rate and SNR
age and environment, there may exist a short window in the pulsar's life when 
DMs are larger than could be achieved in the IGM at redshifts $z\lesssim2.5$. 
Na\"ively we would expect the average pulse 
energy to decay with time along with its period. 
It would be therefore possible, albeit rare, that an FRB have a DM 
 of $\sim10^4$ pc cm$^{-3}$. In general we expect the distribution of DMs 
to be peaked somewhere around the observed FRBs (500-800  pc cm$^{-3}$),
but with weight at intermediate 
DMs when the ejecta has significantly expanded and at very high DMs. In several
non-cosmological FRB models there should be a number of low-DM FRBs
 \citep{2015arXiv150101341P, 2015arXiv150400200X}, which must be explained
 away with non-identification bias. However in our 
picture we do not expect the pulsar to emit supergiant pulses indefinitely and 
therefore we do not expect to be able to see these objects when the SNR has 
expanded and the DMs would be moderate. In the \cite{2015arXiv150506220K} treatment
of the SNR FRB
it is assumed that the supergiant pulse rate is time-independent, a scenario that the 
observed $\frac{dlogN}{dlogDM}$ has already cast doubt on. But since the
DM distribution depends on birth spin rate and the dependence of luminosity 
with period, both of which are unknown, we do not attempt to predict it concretely. 
The Crab would need to emit giant pulses in excess of several GJy to be 
seen at the distances we are proposing - which is several orders of
magnitude brighter than what has been observed - 
and we postulate that is because it is too old.

Beyond the varying DM distributions produced by the location, density, 
and time dependence of the dispersing electrons, their plasma frequency 
can give interesting constraints on the nature of FRBs. Since 

\begin{equation}
\omega_p = \sqrt{\frac{n_e e^2}{m_e \epsilon_o}}
\end{equation}

\noindent we expect FRBs dispersed by 
the diffuse IGM to have very low plasma frequencies while Galactic models,
e.g. flaring stars, should predict large $\omega_p$. This can be verified 
in in precise measurements of the pulse arrival time as a function of frequency.
If $k^2c^2 = \omega^2 - \omega_p^2$, then the $\lambda^2$ arrival time 
dependence is only true in the limit where $\omega >> \omega_p$.
Therefore plasma frequency can be used to test FRB models 
by looking for deviations in the data. This probe 
was also pointed out by \cite{2015arXiv150506220K}, 
who shows the dispersion index for
$\Delta t \propto \nu^{\alpha}$ differs from -2 by 
$\frac{6 \pi n_e e^2}{m_e \omega^2}$.
In the SNR model a $\sim$50 year SNR expanding at
$\sim$ 3,000 km s$^{-1}$
would have a plasma frequency of $\lesssim 10$ MHz and an arrival 
delay within $4 \times 10^{-5}$ of -2.0 at 1.4 GHz, which is consistent with 
present measurements. 
 
Another interesting path for studying extragalactic radio bursts, 
cosmological or otherwise, is scintillation. Only objects of small angular
size scintillate, which is why stars twinkle and planets do not: turbulent cells
in the ionosphere can resolve planets but not stars. The same is true for extragalactic
objects scintillating in the Milky Way, where objects larger than $\sim10^{-7}$ 
arc seconds do not scintillate at $\sim$GHz. 
This is why so few quasars scintillate \citep{2002Natur.415...57D}. 

While several explanations for this scintillation exist 
\citep{1992RSPTA.341..151N, 2014MNRAS.442.3338P}, we are  
concerned with the observational effects and not the physics. 
Using \cite{1986isra.book.....T} 
we can estimate the angular size of an extragalactic object,

\begin{equation}
\label{eqn-scint}
\theta \approx \left ( \frac{2 c \tau \, (R_{\rm obj} - R_{\rm sn})}{R_{\rm sn}R_{\rm obj}} \right )^{1/2}
\end{equation}
\\

\noindent where $R_{\rm obj}$ is the distance to the source, $R_{\rm sn}$ is the distance 
to the screen, and $\tau$ is the scattering timescale. For the case of FRBs we 
take $\tau$ to be $\sim10$ ms. In the cosmological case, if the ms scattering
were from an extended galactic disk along the line of sight  (see \citet{2014ApJ...780L..33M})
halfway between
us and the source, then the angular broadening
of an object at 2 Gpc is $\sim150$ microarcseconds. If the screen were within 
1 kpc of the same object then the broadening is $\sim80$ nanoarcseconds. 
Therefore scintillation from our own Galaxy should only occur for cosmological 
FRBs whose millisecond scattering is close to the source. For an SNR FRB the 
screen would have to be within a few hundred parsecs of the object, which we 
generically expect. We include this feature in table 1 where each 
column is estimated based on the medium that is causing the high dispersion measure, 
e.g. the IGM for cosmological models.

\section{Conclusions}
Evidence is emerging suggesting FRBs are not only extraterrestrial
but extragalactic. Though the simplest interpretation of their high DMs 
is a cosmological one, we find this model less compelling in the light of 
past scattering measurements and potential Faraday rotation 
and pol-angle swing in a new FRB 
(Masui et al. 2015, submitted). In this letter we offer a 
more nearby solution. 
We have gone through
a model in which FRBs are really supergiant pulses from 
extragalactic supernova remnant pulsars, within a couple hundred megaparsecs. 
The SNR environment is sufficiently
dense and ionized to provide DMs $\gtrsim 500$pc cm$^{-3}$ as well as 
RMs $\gtrsim 50$ rad m$^{-2}$, only the first of which could be replicated by the IGM. 

The environment could also provide $\sim$ms scattering at 1 GHz, as has been 
observed in Galactic SNR pulsars. 
That makes this picture self-contained in the sense that
the young remnant environment can account for the dispersion 
and scattering measure seen in FRBs.  It predicts a higher Faraday
rotation than the IGM, but not as high as galactic centers. The
repetition rate is related to the distance, and could be from days to
years. 
By extrapolating Crab-like giant pulses back to the pulsar's first century or so,
we have proposed that such objects can emit extremely energetic bursts sporadically. 
If these are similar to giant pulses from Galactic pulsars, they could be highly polarized, 
either linearly or circularly, and if they were to repeat their polarization state may 
change drastically. Given the object's rotating nature, polarization
 angles would be likely to swing during the pulse. The distinct polarization properties
have been seen in at least one burst and may be end up being generic properties 
of FRBs (Masui et al. 2015, submitted). 
\\

\section{Acknowledgements}

We thank NSERC for support. We also thank Bryan Gaensler, Niels Oppermann, 
Giovanna Zanardo, and Chris Matzner for helpful discussions. 

\newcommand{\araa}{ARA\&A}   
\newcommand{\afz}{Afz}       
\newcommand{\aj}{AJ}         
\newcommand{\azh}{AZh}       
\newcommand{\aaa}{A\&A}      
\newcommand{\aas}{A\&AS}     
\newcommand{\aar}{A\&AR}     
\newcommand{\apj}{ApJ}       
\newcommand{\apjs}{ApJS}     
\newcommand{\apjl}{ApJ}      
\newcommand{\apss}{Ap\&SS}   
\newcommand{\baas}{BAAS}     
\newcommand{\jaa}{JA\&A}     
\newcommand{\mnras}{MNRAS}   
\newcommand{\nat}{Nat}       
\newcommand{\pasj}{PASJ}     
\newcommand{\pasp}{PASP}     
\newcommand{\paspc}{PASPC}   
\newcommand{\qjras}{QJRAS}   
\newcommand{\sci}{Sci}       
\newcommand{\solphys}{Solar Physics}       %
\newcommand{\sova}{SvA}      
\newcommand{\aap}{A\&A}
\newcommand\jcap{{J. Cosmology Astropart. Phys.}}%
\newcommand{\prd}{Phys. Rev. D}

\bibliography{SNRFRB_April30_MNRAS}
\bibliographystyle{mn2e}

\label{lastpage}

\end{document}